\documentstyle{l-aa}
\font\sixrm=cmr6
\def\grad{\hbox{$^\circ$}}
\def\bmin{\hbox{$^\prime$}}
\def\std{\hbox{$^{\rm h}$}}
\def\min{\hbox{$^{\rm m}$}}
\def\bsec{\hbox{$.\!\!{\arcsec}$}}
\def\dsec{\hbox{$.\!\!^{\rm s}$}}
\newcommand{\RA}[4]{#1\std #2\min #3\dsec#4} 
\newcommand{\DEC}[4]{$#1$\grad $#2$\bmin #3\bsec#4}      
\def\magn{\hbox{\hbox{$^{\rm m}$}} }

\newcommand{\magpt}[2]{\mbox{$\rm #1\hspace{-0.25em}\stackrel{m}{.}
      \hspace{-1.0mm}#2$}}                             
\newcommand\ion[2]{\hbox{#1\,{\sixrm #2}}}
\newcommand\Teff{$ T_{\rm eff}$}
\newcommand\teff{$ T_{\rm eff}$}

\newcommand\logg{$\log {\rm g}$}
\newcommand{\Msolar}{\mbox{\,$\rm M_{\odot}$}}        
\begin{document}
\thesaurus{1(08.05.1; 08.19.2; 08.16.3; 10.07.3 M~15)}
\title{Hot HB stars in globular clusters - physical parameters and consequences
for theory}
\subtitle{IV. sdB candidates in M~15}

\author{S. Moehler
\inst{1} \inst{2}
\thanks{Visiting Astronomer, German-Spanish Astronomical Center,
Calar Alto,
operated by the Max-Planck-Institut f\"ur Astronomie jointly with the
Spanish National Commission for Astronomy}
 \and U. Heber \inst{3}
 \and P.R. Durrell \inst{4} }
\offprints{U. Heber}
\institute {Landessternwarte, K\"onigstuhl, 69117 Heidelberg, Federal
Republic of Germany
\and Space Telescope Science Institute, 3700 San Martin Drive, Baltimore,
MD 21218, USA (e-mail: smoehler@stsci.edu)
\and Dr. Remeis-Sternwarte, Sternwartstr. 7, D 96049 Bamberg, Germany
\and Dept. of Physics \& Astronomy, McMaster University, Hamilton, ON,
Canada, L8S 4M1
}
\maketitle
\begin{abstract}
Quantitative spectroscopic analyses of two faint blue stars
(V=\magpt{19}{5} -- \magpt{20}{0}) in the globular cluster
M\,15 are presented. Their derived \Teff, gravities and absolute magnitudes
(\Teff=24000K, log\,g=5.2, M$_{\rm V}$=\magpt{4}{3}; \Teff=36000K, log\,g=5.9,
M$_{\rm V}$=\magpt{4}{7},
respectively) are matched very well by models for the Extreme
Horizontal Branch (EHB). Both stars are bona-fide subdwarf B stars
making M\,15 only the second globular cluster
(after NGC\,6752) for which the existence of sdB stars has been proven
spectroscopically. While the helium abundance (one tenth solar)
of F1-1 is typical
for sdB stars, F2-2 surprisingly turned out to be a helium rich star, the
first to be reported as a member of
a globular cluster. In the field population
of the Milky Way such stars are rare (less than 5\% of all sdB stars).
 From its proximity to the helium main sequence, it is speculated that
F2-2 may be a naked helium core, i.e.
an Extreme Horizontal Branch star which lost (almost) all of its
hydrogen-rich envelope.

\end{abstract}

\thesaurus{1(08.05.1;08.05.2;08.16.1;10.08.1)}

\section{Introduction}

Subluminous B stars (sdB) form the extreme blue end (\teff\,$>$~20000 K and
\logg\,$>$~5) of the Horizontal Branch
(Heber et al., \cite{hehu84}) and are therefore termed Extreme Horizontal
Branch (EHB)
stars. Such objects are now believed to be the dominant source of UV
radiation causing the UV upturn phenomenon in elliptical galaxies and galaxy
bulges (Dorman et al., \cite{dooc95}).
While there are more than a thousand sdBs known in the field of our
galaxy (Kilkenny et al., 1988), only in one globular cluster bona-fide sdB
stars were shown to exist (NGC 6752, Heber et al., \cite{heku86}, Moehler et
al., \cite{mohe96}).
Several claims that sdB stars had been found in other globular
clusters too could  not be confirmed by spectroscopic analyses.
Moehler et al. (\cite{mohe95}, \cite{mohe96}, de Boer et al. \cite{dbsc95},
hereafter Paper I,III, and II respectively)
found that all ``classical''
BHB stars (i.e. stars with \teff\ $<$ 20000~K and \logg\ $<$ 5)
in several clusters
exhibited too low masses compared to standard evolutionary theory, whereas
the sdB stars' masses in NGC~6752 were in good agreement with the canonical
mass of 0.5 \Msolar.

It is therefore of great importance
to find and analyse sdB stars in other globular clusters.
In this letter we present follow-up spectroscopy and spectral
analyses of faint blue
stars (19\magn $<$ V $<$ 20\magn and \magpt{$-$0}{28} $<$ (B$-$V) $<$
\magpt{$-$0}{12})
in the globular cluster M\,15, which were discovered recently by
Durrell \& Harris (\cite{duha93}).

\section{Observations and Reduction}

Two of the four candidates (F2-1 and F2-3)
could not be observed reliably from the ground due to nearby
red neighbours (see Table 1).
The remaining candidate stars were observed with
the focal reducer of the 3.5m telescope at the Calar Alto observatory using
grism \#3 (134~\AA/mm) and a slit width of 1\bsec5,
resulting in medium resolution spectra
covering the wavelength range 3250 - 6350~\AA. We also obtained
low resolution spectrophotometric data, using a slit width of 5\arcsec\ and
binning the spectra by a factor of 2 along the dispersion axis.

\begin{table}
\begin{tabular}{|l|rr|rr|}
\hline
Number & V & B$-$V & $\alpha_{2000}$ & $\delta_{2000}$ \\
\hline
F1-1 & \magpt{19}{468} & \magpt{-0}{160} & \RA{21}{30}{12}{3} &
\DEC{+12}{03}{45}{6}\\
F2-1 & \magpt{19}{101} & \magpt{-0}{126} & \RA{21}{29}{39}{6} &
\DEC{+12}{07}{26}{3}\\
     & \magpt{18}{555} & \magpt{+1}{531} & \RA{21}{29}{39}{6} &
\DEC{+12}{07}{26}{8}\\
F2-2 & \magpt{19}{983} & \magpt{-0}{231} & \RA{21}{29}{34}{7} &
\DEC{+12}{09}{19}{1}\\
F2-3 & \magpt{19}{956} & \magpt{-0}{276} & \RA{21}{29}{46}{6} &
\DEC{+12}{06}{53}{7}\\
     & \magpt{18}{027} & \magpt{+0}{702} & \RA{21}{29}{46}{6} &
\DEC{+12}{06}{54}{3}\\
\hline
\end{tabular}
\caption{Positions and magnitudes of the sdB candidates and their close
neighbours}
\end{table}

Since there is no built-in wavelength calibration lamp we observed wavelength
calibration spectra only at the beginning of the night. The observation
and calibration of bias, dark current and flat-field was performed as
described in paper I and III. As the focal reducer produces rather strong
distortions we extracted that part of each observation that contained the
desired spectrum and some sky and straightened it, using a program written
by Dr. O. Stahl (priv. comm.).
We applied the same procedure to the wavelength
calibration frames. Thereby we could perform a good two-dimensional wavelength
calibration (using the MIDAS Long context)
and sky-subtraction (as described in paper I).
Correction of atmospheric and interstellar extinction as well as flux
calibration was again performed as described in paper I
using the flux tables of Massey (\cite{mast88}) for BD+28$^\circ$4211.

\section{Analyses}

\subsection{F1-1}
The spectra show broad Balmer lines and a weak He~I line at 4471~\AA\
typical for sdB stars.
The low resolution data unfortunately showed some excess flux towards
the red, probably caused by a red star (V = \magpt{19}{99}, B$-$V =
\magpt{+0}{55}) 3\arcsec\ away.
Fitting only the Balmer jump
and the B and V photometry of Durrell \& Harris (1993) we get an effective
temperature of 24000~K (cf. paper III).
We used the helium- and metal-poor model atmospheres of Heber
(cf. Paper III) to analyse the medium resolution spectrum
of F1-1. As the resolution of the spectrum varies with wavelength we convolved
the model atmospheres with different Gaussian profiles for the three Balmer
lines. The appropriate FWHM were determined from the calibration lines close
to the position of the Balmer lines. We used FWHM of 8.7 \AA\ for H$_\delta$,
7.1 \AA\ for H$_\gamma$, and 9.8 \AA\ for H$_\beta$. We fitted each line
separately and derived the mean surface gravity
by calculating the weighted mean of the individual results. The weights
were derived from the fit residuals of the individual lines. We thus get
a log g value of 5.2$\pm$0.14 for a fixed effective temperature of 24000~K.
The fits to the Balmer lines are shown in Fig. 1. Note in passing
that higher temperature and gravity results if we ignore the Balmer jump
and derive \Teff\ and \logg\ simultaneously from the three Balmer line
profiles alone:
The smallest overall residual is achieved for
\Teff\ = 29700~K and \logg\ = 5.9. These values, however,
are inconsistent with the low resolution data.
We already noted similar inconsistencies
between Balmer jump and Balmer line profiles in paper I, which are probably
caused by insufficient S/N in the Balmer line profiles.
The \ion{He}{I} 4471~\AA\ line is consistent with a helium abundance of
about 0.1 solar.
Using the routines of R. Saffer (Saffer et al., \cite{saff94})
and a fixed mean FWHM of 8~\AA\ we get internal errors
of 600 K and 0.13 dex, respectively. We take the standard deviation of \logg\
as external errors  and assume an external error of \teff\ of $\pm$ 2000~K
(cf. Paper~III).
Using the same method as in Papers I and III we get a logarithmic mass of
$-$0.423$\pm$0.20 dex, corresponding to (0.38$^{+0.22}_{-0.14}$)~\Msolar.
For a detailed discussion of the errors entering into the mass determination
see Paper~III.

\begin{figure}
\vspace{6cm}
\includegraphics{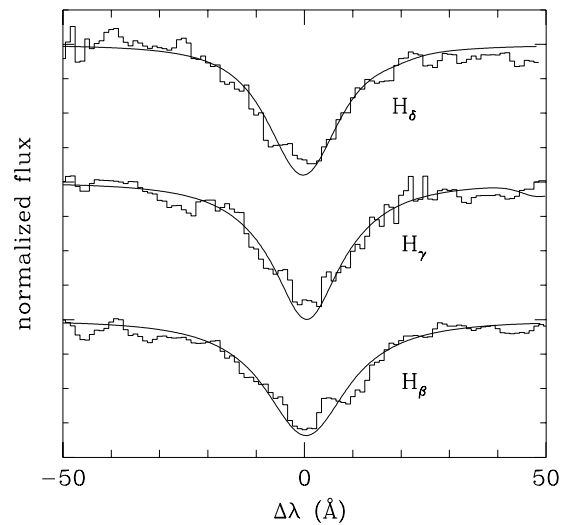}
\caption[]{The best fitting models for \teff\ = 24000~K
compared to the Balmer lines of of F1-1 (\logg\ = 5.0 (H$_\delta$),
5.3 (H$_\gamma$, H$_\beta$)).
The tickmarks mark steps of 10\% in intensity.}
\end{figure}

\subsection{F2-2}

Unlike F1-1 and other sdBs the spectrum of F2-2 is not dominated by Balmer
lines but displays a prominent \ion{He}{I} absorption line spectrum in
addition to the Balmer lines (see Fig.2). The Balmer lines are much weaker
and narrower than in F1-1.
In Fig. 2 we compare the spectrum of F2-2 (smoothed by a box filter
of 5 \AA\ width) to that of a
He-sdB in the field (HS\,0315+3355, Heber et al. \cite{hedr96}), a spectral
type
that describes rare helium-rich variants of the sdB stars
(Moehler et al., \cite{mori90}, see also below),
The helium line spectra of both stars are
very similar, while the Balmer lines in HS\,0315+3355 are considerably weaker
than in F2-2 (see Fig. 2), indicating an even lower hydrogen content of the
former.

\begin{figure}
\vspace{6.0cm}
\includegraphics{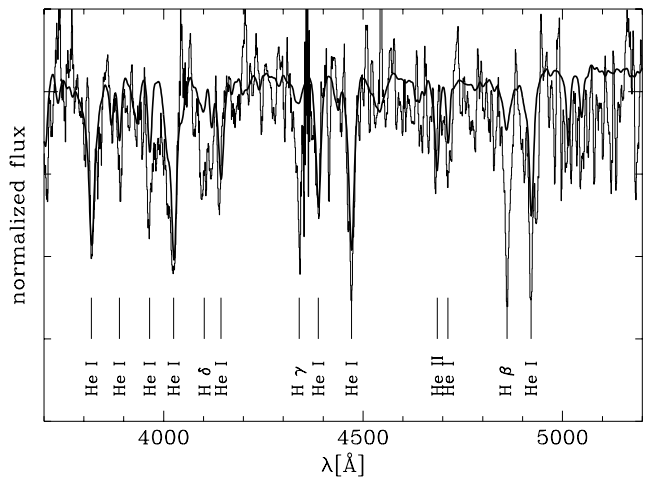}
\caption[]{The medium resolved spectrum of F2-2 and the identified absorption
lines. To allow a better identification of the lines the spectrum has been
smoothed with a box filter of 5\AA\  width.
In comparison (thick line)
we show the spectrum of a helium-rich sdB star (HS\,0315+3355) in the field
that has been convolved to the same resolution. The tickmarks mark steps of
10\%
in intensity.}
\end{figure}

The admittedly
rather low S/N unfortunately allows only a coarse analysis. Since an
analysis of individual spectral lines is impossible
we choose to fit
the entire useful portion of the spectrum (4000\AA\ -- 5000\AA)
simultaneously. It turned out
that F2-2 is somewhat hotter than F1-1 and we therefore used the updated
grid of NLTE model atmospheres calculated by S. Dreizler (see Dreizler et
al. \cite{drei90}).
The model atmospheres include detailed H and He model atoms and the
blanketing effects of their spectral lines but no metals. Using Saffer's
fitting program \teff , \logg , and helium abundance were determined
simultaneously
from a small subgrid (\teff\ = 35, 40, 45kK; \logg\ = 5.5, 6.0, 6.5;
He/(H+He)~=~0.5, 0.75, 0.91, 0.99, by number).

Since the \ion{He}{I} 4144\AA\ line is not included in the models, this
line is excluded from the fit.

\begin{figure}
\vspace{5.7cm}
\includegraphics{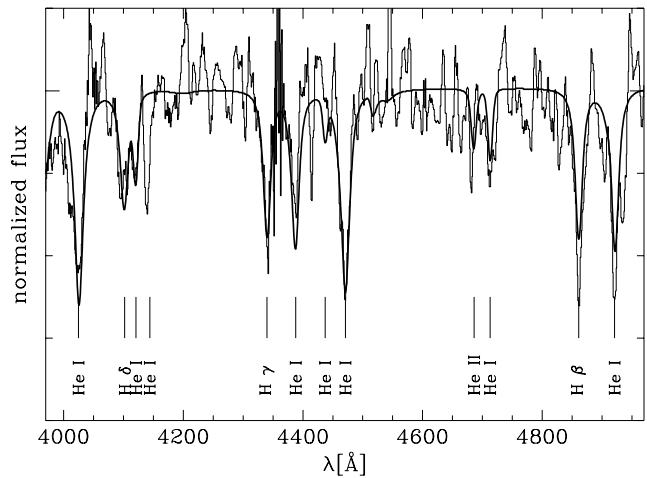}
\caption[]{The spectrum of F2-2 and the best fitting model
(thick line, \teff\,=~36000~K, \logg\,=~5.9,
N$_{He}$/(N$_{H}$+N$_{He}$)~=~0.87).
To allow a better identification of the lines the observed spectrum has been
smoothed with a box filter of 5\AA\  width. The tickmarks mark steps of 10\%
in intensity.}
\end{figure}

An effective temperature of 36000~K, \logg\ of 5.9, and
a helium abundance of N$_{He}$/(N$_{H}$+N$_{He}$)~=~0.87
(by number) resulted. The fit to the
spectrum is displayed in Fig. 3.
Since the noise level is rather high it is
difficult to estimate the error ranges of the atmospheric parameters.
As a test we changed the fitted
spectral range as well as the placement of the continuum, which resulted
in quite small changes of \teff\ ($\approx$ 2000K) and \logg\ ($\approx$
0.2~dex).
The helium abundance N$_{He}$/(N$_{H}$+N$_{He}$) ranged from 0.5 to 0.9 by
number, indicating that helium is overabundant by at least a factor 5 with
respect to the sun. As conservative error
estimates we adopted $\pm$ 4000~K and 0.5 dex for the errors in \teff\ and
\logg .

\subsection{Radial velocities}

For F1-1 we derive a mean heliocentric velocity of $-$142~km/sec
from the Balmer lines with an estimated error of about $\pm$40~km/sec
(due to the mediocre
resolution, the limited accuracy of the wavelength calibration,
and the shallow lines). This value is within the
error limits consistent with the cluster velocity
of -107.09 km/sec (Peterson et al., \cite{pese89}).
For F2-2  we could not derive reliable radial
velocities from single lines due to the low S/N of the spectrum.
Instead we cross correlated the normalized
spectrum with that of the field He-sdB HS\,0315+3355
(convolved to the same resolution).
This procedure resulted in a heliocentric velocity of $\approx -$70~km/sec,
which - due to the large errors - does not contradict a cluster membership
for F2-2.

\section{Discussion}

\begin{figure}[t]
\vspace{6.5cm}
\includegraphics{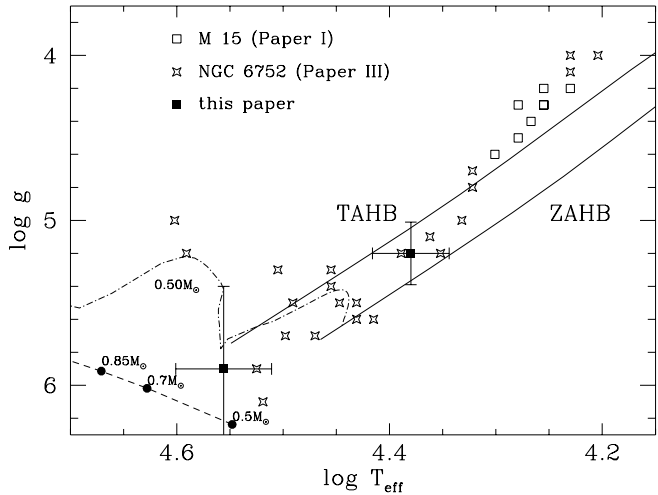}
\caption[]{The physical parameters of the sdBs in M15 compared to
the results of Papers I and III and to theoretical expectations.
The solid lines are the Zero Age HB and the Terminal Age (core
helium exhaustion) HB for [Fe/H]~=~$-$2.26 of Dorman
et al. (\cite{dorm93}). The short dashed line gives the position of the
helium main sequence (Paczynski 1971). The long dashed-short dashed lines
give post-EHB evolutionary tracks by Dorman et al. (\cite{dorm93}), labeled
with
the total mass of the EHB star.}
\end{figure}

The spectroscopic analyses of two faint blue stars in the globular cluster
M\,15 show that both stars are bona-fide subdwarf B stars. In Fig. 4 and
5 we compare their positions in the (\Teff, log\,g)- and the
(\Teff, M$_{\rm V}$)-diagrams to those of EHB stars in NGC\,6752 (from
paper III) as well as to the predictions from
Horizontal Branch stellar models (from Dorman et al. \cite{dorm93}).
As can be seen from this comparison,
their evolutionary status is well described by models for the Extreme
Horizontal Branch (EHB). Hence M\,15 is only the second globular cluster
(after NGC\,6752) for which the existence of EHB stars has been proven
spectroscopically. While the helium abundance of F1-1 is typical
for sdB stars (i.e. subsolar),
F2-2 surprisingly turned out to be a helium rich star.

This is the first time ever that a helium rich sdB star
has been reported in a globular cluster. In the field
of the Milky Way only 5\% of the sdB stars are helium-rich. Jeffery et
al. (\cite{jehe96}) list 48 such stars, while the latest (unpublished)
version of the catalog of hot subdwarfs (Kilkenny et al., \cite{kihe88})
lists more than 1,000 hydrogen-rich sdBs.

The helium-rich sdB has an absolute visual brightness of about
\magpt{4}{7}, which places it at the very faint blue end of the EHB as seen
in the
colour-magnitude diagram of NGC 6752. F2-2 may even be hotter than
any EHB star in NGC\,6752. From its proximity to the helium main sequence in
Fig.\,4 and 5 it might be tempting to regard F2-2 as a naked helium core, i.e.
as an Extreme Horizontal Branch star which lost (almost) all of its
hydrogen-rich envelope.

Why didn't we find any helium-rich sdBs in NGC\,6752?
All the EHB stars that have
been analysed in NGC 6752 (including the three faintest ones seen in Buonanno
et al., \cite{buca86}) are helium-poor sdB stars (Paper III). This could
either mean
that there are no helium-rich subdwarfs in NGC~6752 or that they are just
below the detection limit of Buonanno et al. (\cite{buca86}). One should
certainly keep an eye on newer and deeper CMDs of globular clusters to
see whether other He-sdB candidates show up.

\begin{figure}
\vspace{6.cm}
\includegraphics{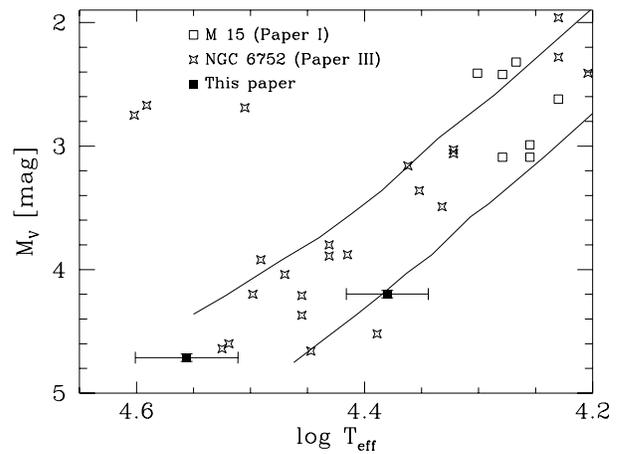}
\caption[]{The absolute V magnitudes and effective temperatures
as given above compared to theoretical tracks by
Dorman et al. (\cite{dorm93}, details see Fig.~4).
Also shown are the data for the stars analysed in papers I and III.}
\end{figure}

\acknowledgements
We thank Dr. S. Dreizler for making available his NLTE model grid
to us and the staff of the Calar Alto Observatory for their support
during the observations. Thanks go also to Dr. R. Saffer for valuable
discussions. SM acknowledges support from the DFG (Mo 602/5-1),
by the Alexander von Humboldt-Foundation, and
by the director of STScI, Dr. R. Williams, through a DDRF
grant.

{}
\end{document}